\documentclass[12pt]{article}

 \newif\ifpdf
 \ifx\pdfoutput\undefined
 \pdffalse 
 \else
 \pdfoutput=1 
 \pdftrue
 \fi

\ifpdf
    \pdfcatalog{ /PageMode (/UseNone)
                 /OpenAction (fitbh)
                }
   \usepackage[pdftex,pdftitle={},urlcolor=blue,colorlinks=true]{hyperref}
   \usepackage[pdftex]{graphicx}
\else
   \usepackage[ps2pdf,pdftitle={PDF},urlcolor=blue,colorlinks=true]{hyperref}
   \usepackage{graphicx}
\fi

\newcommand{\nin}{\noindent}

\newcommand{\be}{\begin{equation}}
\newcommand{\ee}{\end{equation}}
\newcommand{\bea}{\begin{eqnarray}}
\newcommand{\eea}{\end{eqnarray}}

\newcommand{\nonum}{\nonumber}
\begin{document}

\ifpdf
\DeclareGraphicsExtensions{.pdf, .jpg, .tif}
\else
\DeclareGraphicsExtensions{.eps}
\fi
\nin Martina Hentschel and Jens U.~N{\"o}ckel\footnote{\href{http://darkwing.uoregon.edu/~noeckel}{\em Current address}: Department of Physics,
    University of Oregon, 1371 E 13th Avenue, Eugene, OR 97403}\\
\bigskip
\parbox[t]{12.5cm}{
{\footnotesize
{\em Published in} Quantum Optics of Small Structures, edited by D.Lenstra, T.D.Visser and K.A.H.van Leeuwen (Edita KNAW, Amsterdam, 2000)
}}
\bigskip
\medskip

\nin{\bf\Large  
The sequential-reflection 
model in\\ 
deformed dielectric cavities}
\bigskip

\nin
\parbox[t]{9cm}{
\nin
{The stationary states of a microlaser are related to the 
decaying quasibound states of the corresponding passive cavity. These
are interpreted classically as originating from sequential escape
attempts of an ensemble of rays obeying a curvature-corrected Fresnel
formula. Polarization-dependent predictions of this model, and its
limitations for stable orbits in partially chaotic systems are 
discussed.
}
}
\bigskip

\nin 
As a mechanism for achieving mode confinement, waveguiding by total 
internal reflection is ubiquitous in optics. However, in dielectric 
microresonators where three-dimensionally confined mode volumes are 
desired, there is always leakage because the ray 
picture, in which Fresnel's formulas describe the outcoupling, 
acquires corrections. Leaky modes corresponding to classically confined 
rays can be found, e.g., in optical fibers as ``spiral'' 
modes (Poon), or in latterally structured cylindrical 
VCSELs (Ahn) as well as in microdisk lasers (McCall). The 
classically forbidden loss in such modes is analogous to tunneling 
through an effective potential barrier (Johnson). 

The highest Q is achieved for modes which semiclassically correspond 
to rays almost at grazing incidence. Resonators with a circular 
cross section are a particularly simple realization of this requirement, 
because they exhibit {\em whispering-gallery} (WG) modes characterized 
by high intensity in an annular region near the surface. 
However, even Lord Rayleigh who 
first described the acoustic analog that gave the phenomenon its name,
concluded (Strutt) that it requires only an everywhere positive 
curvature, not necessarily rotational symmetry. A rigorous proof of
this is difficult because in the short-wavelength 
limit, this ``clinging'' of waves to the walls has to carry over to 
the ray picture, in which a generic oval
cavity exhibits a transition to chaos (Lazutkin; N{\"o}ckel, 1996a). 
Notwithstanding, this problem is fundamental to microresonator 
design (N{\"o}ckel, 1997a), because the 
availability of high-Q modes is the foremost selection criterium
in an otherwise unbounded space of potential resonator 
shapes (Angelow). 
Chaos can in fact make WG modes more useful, and moreover create 
other types of modes with desirable properties, such as the bowtie 
pattern whose confocal geometry 
points the way toward the strong-coupling regime in combination with 
focused emission (Gmachl, Morin). 

The robustness of whispering-gallery type intensity patterns in the
modes of convex resonators extends even to
nonlinear media (Harayama). However, in that case the
distinction to the widely studied phenomenon of vortex formation
(Weiss) becomes washed out: a WG mode is also a vortex with a
phase singularity at points of vanishing intensity; for 
a circular resonator where the field is proportional to a Bessel
function $J_m(kr)\approx r^m$ near the center $r= 0$, the
vorticity is simply the angular momentum quantum number $m$. 
Therefore, since our aim is to address 
the fundamental aspect of the shape dependence of high-Q modes in 
microresonators, we focus here on linear media where amplification 
is taken into account by a negative imaginary part of the refractive
index ${\tilde n}$. 

The model considered here can be derived
from a homogenous cylinder by deforming its cross section and
considering only propagation transverse to its axis. In this case, 
TE and TM polarization are decoupled and one has to consider only a 
scalar wave equation 
\be\label{eq:wavescalar}
\nabla^2\psi+{\tilde n}^2k^2\,\psi =0,
\ee
assuming a steady state time dependence so that $k$ is real. Here,
${\tilde n}\equiv n - i\,n'$ inside the resonator and ${\tilde n}=1$ 
outside, giving 
rise to an exterior and interior field, $\psi_{\rm ext}$ and  $\psi_{\rm
int}$, both of which are connected by the proper matching conditions
at the dielectric interface, depending on polarization. For TM modes, 
$\psi$ denotes the electric field, which is parallel to the cylinder
axis. In this case one finds that $\psi$ and its normal derivative are 
continuous at the interface, in analogy to quantum mechanics. 

The system is open because it radiates energy into the environment via 
its modal losses. This openness increases as 
$n\to 1$, and the closed-resonator limit is approached for
$n\to\infty$. This can be understood from Fresnel's formulas
which imply total internal reflection for all angles of incidence $\chi$
satisfying $\sin\chi>1/n\equiv\sin\chi_c$ ($\chi_c$ is the {\em critical angle}).
Equation (\ref{eq:wavescalar}) can be recast as
\be\label{eq:wavescalarqb}
\nabla^2\psi+n^2{\tilde k}^2\,\psi =0,
\ee
where $n$ is the real part of ${\tilde n}$ as defined above, and 
${\tilde k}\equiv k-i\,k\,n'/n$ is a complex wavenumber inside the 
cavity but reduces to ${\tilde k}= k$ outside. If instead of this we 
also had ${\tilde k}= k-i\,k\,n'/n$ outside, the solutions of 
Eq.\ (\ref{eq:wavescalarqb}) would be the {\em quasibound states} of the 
passive resonator, as they arise when one assumes 
a decaying time dependence $\propto \exp[-ick\,t-\gamma\,t]$, where 
$\gamma=ck\,n'/n$. 

For a quasibound (or metastable) state, the field at distances larger
than $\approx c/(2\gamma)$ from the cavity grows exponentially due to
retardation, but within this physical range $\psi_{\rm ext}$ vanishes 
as $\gamma\to 0$, so that one can write $\psi_{\rm ext}({\bf r})\approx \gamma \zeta({\bf r})$. 
If one expands
the dependence of $\psi_{\rm int}$ and $\zeta$ on $\gamma$ in a Taylor series,
then to linear order the  $\gamma$-dependence of $\zeta$, but not that of 
$\psi_{\rm int}$, can be dropped in the full solution. Therefore, the 
stationary state of the active medium and the metastable decaying 
state are identical to first order in $\gamma$ within an area of order $\gamma^{-2}$. 

This approximate equivalence establishes a connection to the study of
S-matrix poles from which quasibound states arise, see (Schomerus), and to 
dissipation in quantum mechanics (Prigogine, Braun). The recent
resurgence of interest in these problems is motivated to a significant 
extent by our lack of understanding of the quantum-to-classical
transition, in particular in the presence of classical chaos. 
Precisely this constellation is also present in 
Eq.\ (\ref{eq:wavescalarqb}) when one considers its short-wavelength
limit for the generic case of a deformed cavity. 

In the context of laser resonators, there are 
three main differences to previous work on open quantum systems: 
firstly, we are interested in the properties of {\em individual} 
states of an open system, as opposed to a statistical ensemble; 
see also (Casati). Secondly, an important quantity 
that can be studied for such individual states is their {\em emission 
directionality}, which in other open systems of chemical or 
nuclear physics is averaged out. Finally, the classical limits of
quantum mechanics with smooth potentials and optics with discontinuous 
refractive indices are qualitatively different (Kohler): 
the first yields
deterministic Hamiltonian mechanics; the second leads to the
probabilistic Fresnel formulas which moreover depend on polarization. 

In principle, Eq.\ (\ref{eq:wavescalarqb}) can 
be solved numerically to find the discrete complex ${\tilde
k}$ and the corresponding modes. One approach is based on
the Rayleigh hypothesis (van den Berg, Barton) which in our
implementation for quasibound states (N{\"o}ckel, 1996a) 
assumes that the fields can be expanded in cylinder functions as 
\bea\label{psiintdecompgeneqn}
\psi_{\rm int}(r,\phi)&=&\sum\limits_m A_m\,J_m(kr)\,e^{im\,\phi},\\
\psi_{\rm ext}(r,\phi)&=&\sum\limits_m B_m\,H_m^{(1)}(kr)\,e^{im\,\phi}.\nonum
\eea
where a polar coordinate system with suitably chosen origin is
used. These expansions always work inside some circle
of convergence for $\psi_{\rm int}$ and outside some other circle for
$\psi_{\rm ext}$, and for a large range of resonator shapes both 
convergence domains contain the dielectric interface where 
the matching conditions are imposed to obtain equations for the
unknown coefficients $A_m$ and $B_m$.

Computational cost can be high here, especially at short
wavelengths, and hence a semiclassical approximation can 
lead to simplifications while preserving physical insight.
The ray picture is a cornerstone of 
classical optics, but its value in the study of open resonators only 
unfolds when the ray dynamics is studied in {\em phase space}
(N{\"o}ckel, 1994, 1996a, 1996b; Mekis), because
Fresnel's formulas determine escape probabilities according to the 
angle of incidence $\chi$, not the position of impact. One can
make use of the physical information contained in this picture in two
ways: Either one starts from Eq.\ (\ref{eq:wavescalarqb}) and takes a
short-wavelength limit (Narimanov); or alternatively, one starts
from the classical dynamics and makes {\em classical} 
approximations that allow one to impose simple quantization
conditions and thus make the connection to the resonator
modes (N{\"o}ckel, 1997a). The question whether these different routes 
meet ``in the middle'' is not straightforward because the
problem of semiclassical quantization in a generic deformed resonator
is not completely solved as yet, owing to the coexistence of both
regular and chaotic motion in their classical phase space. 

Among the advantages of the ray-based approach (N{\"o}ckel, 1997a) are 
its flexibility and computational ease. However, in order for the 
prescription outlined in Ref.\ (N{\"o}ckel, 1997a) 
to correctly describe the limiting 
case of a circular cylinder, one must include the tunneling which in 
the circle is the only loss mechanism. This can be done in the ray
picture with a curvature- and wavelength dependent ``rounding'' of 
Fresnel's formulas which the simulation uses at each reflection along
a ray path. The idea used in (N{\"o}ckel, 1997a) was to
interpret the resonance widths of a circular cylinder in terms of a
``sequential-tunneling'' ansatz: if the intensity of a quasibound
state decays as $\exp[-2\gamma\,t]$, this can be interpreted in the ray
picture as the result of $\nu$ sequential escape attempts with
reflection probability $p_0$, where $\nu$ is the number 
of reflections the ray undergoes during the time $t$. In a circle
of radius $R$, a trajectory characterized by the angle of incidence
$\chi$ has $\nu=ct/(nL)$ reflections during $t$ ($c/n$ is the 
speed of light in the passive medium and $L=2R\,\cos\chi$ is the
geometric path length between reflections). Therefore, one expects 
a decay law $\propto p_0^{\nu}=\exp[ct\,\ln p_0/(nL)]$. 
Comparison with the wave result yields 
\begin{equation}\label{reflprobcirceq}
p_0=\exp(-2\,nL \gamma/c)
\end{equation}
An analytic approximation for $\gamma$ in the circle with TM polarization
has been derived in (N{\"o}ckel, 1997b),
\bea\label{widthrepeat}
\gamma&\approx&-\frac{c}{2nR}\ln\left[\frac{n-1}{n+1}\right]\times\\
&&\frac{2}
{\pi\,kR\,[J_m^2(kR)+Y_m^2(kR)]}.\nonum
\eea
Using the semiclassical expression 
\be\label{eq:sinchidef}
m=nkR\,\sin\chi
\ee
for the angular momentum (N{\"o}ckel, 1996a), one then
obtains the reflecticity in terms of purely classical variables,
$p_0(kR,\sin\chi)$. It reduces to Fresnel's formula in the limit of
large radius of curvature $R$, and by construction reproduces the
width of a mode in the circle if applied locally at each reflection in 
our classical ray model. The latter does not hold uniformly for 
a similar correction derived in (Snyder). 

From the classical limit, it follows that there exists an 
{\em upper bound} on
resonance widths for dielectric cavities with stepped index profiles 
and ``quantum-mechanical'' continuity conditions on $\psi$, because 
the reflectivity $p_0$, 
(for polarization perpendicular to the plane of incidence), is 
bounded away from zero. This minimum $p_{0,min}$ will limit the width
of resonances in a cavity of characteristic size $l$ to $\gamma_{max}=-c\,\ln
p_{0,min}/(2nl)$. Smooth index profiles can also appear discontinuous
on the scale of the wavelength but are eventually resolved as $k\to\infty$, 
allowing arbitrarily small reflectivities at perpendicular incidence. 

\begin{figure}[bt]
\begin{center}
\includegraphics{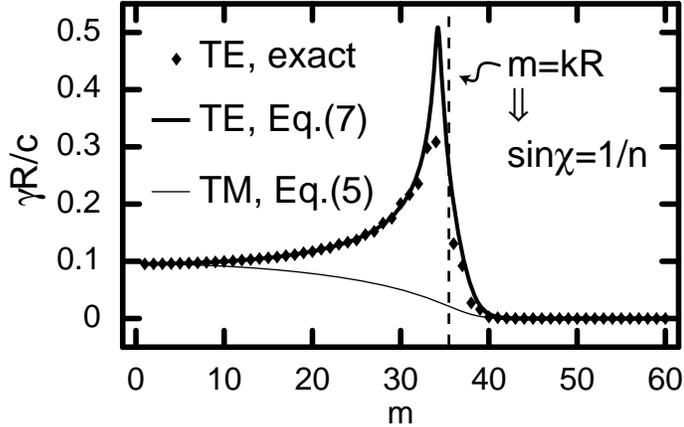}
\end{center}
\vspace*{-0.8cm}
\caption{\label{fig:circle}
Exact resonance widths of a circle (radius $R$, refractive index
$n=3.29$) at $35<kR<35.5$ versus 
angular momentum $m$. Dashed line indicates $\chi_c$ in 
Fresnel's law, using 
Eq.\ (\ref{eq:sinchidef}). Also shown are the TM widths. 
The Brewster angle is at $m\approx 34$ and causes 
a peak in the TE widths.
}\end{figure}

However, extending these arguments to {\em TE polarization} where the
electric field is in the plane of incidence, we furthermore conclude
that a similar upper bound on the widths does {\em not exist} even for 
sharp interfaces. The
reason is that Fresnel's formula yields zero reflectivity at the
{\em Brewster angle} $\chi_B$ at $\sin\chi_B=(1+n^2)^{-1/2}$. The normal derivative of 
$\psi$ (which now represents the magnetic field) exhibits 
a jump proportional to $n^2$ at the dielectric interface -- reminding us 
that this is a situation unique to optics. These general
considerations have important implications for microresonator
design especially at the large $n$ typical for semicondutors, because
in that case $\sin\chi_B\to1/n$, i.e., the ``hole'' in the reflectivity for 
TE polarization approaches $\chi_c$ for total internal
reflection from below. Taking tunneling due to finite curvature into
account as in Eq.\ (\ref{reflprobcirceq}), the rounded Fresnel formula 
then exhibits reduced reflectivity even for
incidence somewhat above $\chi_c$. One can approximately 
obtain the TE widths of the circle from
$\gamma=-c\,{\rm Im}[\sigma] /(nR)$, 
\be
\sigma \approx\arctan[\left(
 n
\frac{H_{m-1}^{(1)}(kR)}{H_{m}^{(1)}(kR)} -
\frac{m}{kR}\,(n-1/n)\right)^{-1}]
\label{eq:widthte}
\ee
This is illustrated in 
Fig.\ \ref{fig:circle}, where refractive index and $kR$ are chosen 
close to those of Ref.\ (Gmachl). The reason is that 
the {\em quantum-cascade} material used there emits 
preferentially TM polarization, whereas the pioneering 
MQW microdisk lasers with sub-micron thickness permit guiding in 
the vertical direction only for TE modes 
[cf. McCall (1992); there, TE/TM must be interchanged to get 
from the slab-waveguide to our cylinder convention]. It is thus 
important to ascertain whether the identical oval lateral design
of the quantum-cascade lasers in Ref.\ (Gmachl) would also 
permit a microdisk laser to operate in TE polarization.

The lasing mode in Fig.\ 3D of (Gmachl) 
was identified as a bowtie-shaped pattern corresponding 
to a periodic ray path with angle of incidence given by 
$\sin\chi\approx1/n$, i.e., directly at the critical angle.
That this mode provides high Q can be seen by comparing to 
$\gamma$ in Fig.\ \ref{fig:circle}: assuming that the width 
$\gamma_{B}$ of a bowtie mode results from the sequential application 
of $p_{0}$ as determined for the circle, the argument leading to 
Eq.\ (\ref{reflprobcirceq}) implies that 
$\gamma_{B}\approx \gamma\,L/l$ where $L/l\approx 1.13$ is 
the ratio of the classical path lengths between reflections in the 
WG orbit and bowtie, respectively. One sees that the TM line 
intersects the critical angle (corresponding to $m=kR$) at a much 
smaller width than the TE curve, and this Q-spoiling due to Brewster 
transmission is borne out by the actual TE resonances as well.
This leads to the prediction that conventional microdisk lasers 
with a shape designed to yield a bowtie pattern just at $\chi_c$ 
as in Fig.\ 3D of (Gmachl) will {\em not lase}.

These ray arguments are known to yield large deviations 
from the true resonance widths when the modes under consideration 
are quantized on stable phase-space domains in a partially chaotic 
system, cf.\ Ref.\ (N{\"o}ckel, 1997a) where this was attributed to 
chaos-assisted tunneling. The latter yields enhanced outcoupling 
and hence the true widths are {\em underestimated} by the 
sequential ray picture. Therefore, the above Q-spoiling for TE modes 
is not counteracted by a correction of this nature. 
The prediction of an upper bound for TM widths is also not affected by 
chaos-assisted tunneling because it cannot be faster than the 
fastest classical process, which in turn is limited by $p_{0}$ at 
$\sin\chi=0$. Beyond this, however, quantitative widths for 
stable-orbit modes in mixed phase spaces are not provided by the ray 
model.

\begin{figure}[bt]
\begin{center}
\includegraphics{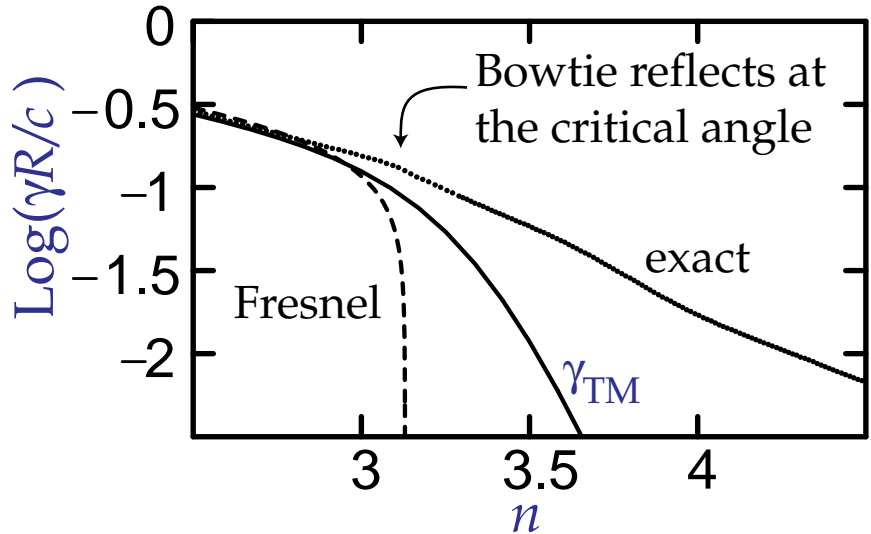}
\end{center}
\vspace*{-0.8cm}
\caption{\label{fig:bowtie}
Width of a TM bowtie mode vs. refractive index, from numerical  
and ray calculations.}
\end{figure}

The disagreement is 
illustrated in Fig.\ \ref{fig:bowtie} for a bowtie mode similar to 
the ones studied in (Narimanov), as a function of $n$, 
but at a deformation of 
$\epsilon=0.16$ [defined as in (Gmachl)] and $nkR\approx 
119.8$. Since $nk$ is the wavenumber
inside the resonator, it should remain approximately independent of
$n$ as long as the outcoupling can be taken into account in the form of a
boundary phase shift intermediate between Dirichlet and Neumann. 
Indeed, for the state shown in Fig.\ \ref{fig:bowtie}, the change in
$nkR$ in the plotted range of $n$ is only $\approx 0.2$. 
The length scale $R$ here is the radius of curvature at the points of 
reflection. 
At small $n$ where $\chi_c$ is larger than the angle of incidence of the
bowtie, escape is classically allowed in Fresnel's formula and hence
curvature corrections are unimportant. At $n>3$, the tunneling
correction in Eq.\ (\ref{widthrepeat}) does improve on the classical
Fresnel prediction ($\gamma=0$) but clearly still underestimates the true
width. As tunneling in general is definable only with respect to a
classical expectation, we could again label the discrepancy as 
chaos-assisted tunneling. However, a semiclassical theory starting
from Eq.\ (\ref{eq:wavescalarqb}) which reproduces the exact behavior
in Fig.\ \ref{fig:bowtie} very well (Narimanov) can shed more light 
on the physics of the phenomenon. 
\bigskip

\nin{\bf \large References:}
\begin{list}{}{\labelwidth1.6cm \leftmargin1.7cm \labelsep0cm
\rightmargin0cm \itemsep0ex \parsep0ex \topsep0ex}
\item[Ahn, J.~C., 1999]{ {\em et al.}, 
Phys.~Rev.~Lett.~{\bf 82}, 536}
\item[Angelow, G., 1996]{, F.~Laeri, T.~Tschudi, Opt.~Lett.~{\bf 21}, 1324
}
\item[Barton, J.~P., 1997]{, Appl.~Opt.~{\bf 36}, 1312}
\item[Braun, D., 1999]{, P.~L.~Braun and F.~Haake, Physica D {\bf 131}, 265}
\item[Casati, G., 1997]{, G.~Maspero and D.~L.~Shepelyanski, 
Phys.~Rev.~E {\bf 56}, R6233
\item[Casati, G., 1999], G.~Maspero and D.~L.~Shepelyanski, 
Physica D {\bf 131}, 311}
\item[Gmachl, C., 1998]{ {\em et al.}, Science {\bf 280}, 1556}
\item[Harayama, T., 1999]{, P.~Davis and K.~S.~Ikeda, Phys.~Rev.~Lett.~{\bf 82}, 3803}
\item[Johnson, B.~R., 1993]{, J.~Opt.~Soc.~Am.~{\bf 10}, 343}
\item[Kohler, A., 1998]{, R.~Bl{\"u}mel, Ann.~Phys.~{\bf 267}, 249}
\item[Lazutkin, V.~F., 1993]{, {\em KAM Theory and Semiclassical Approximations 
to Eigenfunctions}, (Springer, New York)}
\item[McCall, S.~L., 1992]{, A.~F.~Levi, R.~E.~Slusher,
S.~J.~Pearton and R.~A.~Logan, Appl. Phys. Lett. {\bf 60}, 289}
\item[Mekis, A., 1995]{, J.~U.~N{\"o}ckel, 
G.~Chen, A.~D.~Stone and R.~K.~Chang, Phys.~Rev.~Lett.~{\bf 75}, 2682}
\item[Morin, S.~E., 1994]{, C.~C.~Yu and T.~W.~Mossberg, Phys.~Rev.~Lett.~{\bf 73}, 1489}
\item[Narimanov, E.~E., 1999]{, G.~Hackenbroich, P.~Jaquod and A.~Douglas Stone, 
cond-mat/9907109}
\item[N{\"o}ckel, J.~U., 1994]{, A.~D.~Stone and R.~K.~Chang, 
Optics Letters {\bf 19}, 1693}
\item[N{\"o}ckel, J.~U., 1996a ]{, and A.~D.~Stone, in: {\em Optical Processes in
Microcavities}, edited by R.~K.~Chang and A.~J.~Campillo (World
Scientific, Singapore)}
\item[N{\"o}ckel, J.~U., 1996b]{, A.D.Stone, G.Chen, H.~Grossman
  and R.~K.~Chang, Opt.~Lett.~{\bf 21}, 1609}
\item[N{\"o}ckel, J.~U., 1997a]{, A.~D.~Stone, Nature {\bf
    385}, 45}
\item[N{\"o}ckel, J.~U., 1997b]{, Dissertation, Yale University}
\item[Patra, M., 1999]{, H.~Schomerus, C.~W.~J.~Beenakker, cond-mat/9905019}
\item[Poon, A.~W., 1998]{, R.~K.~Chang and J.~A.~Lock, Opt.~Lett.~{\bf 23}, 1105}
\item[Prigogine, I., 1992]{, 
Phys.~Rep.~{\bf 219}, 93}
\item[Snyder, A.~W., 1975]{, J.D.Love, IEEE Trans.~{\bf MTT-23}, 134}
\item[Strutt, J.~W., 1945]{, and Baron Rayleigh, {\em The Theory of Sound}, Vol.II
(Dover, New York)}
\item[Weiss, C.~O., ]{ {\em et al.}, Appl.~Phys.~B {\bf 68}, 151 (1999)}
\item[van den Berg, P.~M., 1979]{, J.T.Fokkema, IEEE 
Trans.~Anten.~Propag. {\bf AP-27}, 577}
\end{list}
\bigskip

\nin
{\it Max-Planck-Institut f{\"u}r Physik komplexer Systeme,
N{\"o}thnitzer Str. 38, 01187 Dresden, Germany}

\end{document}